\documentclass[12pt,letterpaper]{article}
\pdfoutput=1

\usepackage{graphicx,array}
\usepackage{url}
\usepackage{color}
\usepackage{latexsym}
\usepackage{amsthm}
\usepackage{amsmath}
\usepackage{amssymb}
\usepackage{amsfonts}
\usepackage[numbers,sort&compress]{natbib}
\usepackage{bm}
\usepackage{slashed}
\usepackage{mathrsfs}
\usepackage{enumerate}
\usepackage{hyperref} 
\hypersetup{
    colorlinks=true,       
    linkcolor=red,          
    citecolor=blue,        
    filecolor=magenta,      
    urlcolor=blue           
}
\usepackage[all]{hypcap} 

\usepackage{natbib}
\usepackage[normalem]{ulem} 

\newcommand{\nc}{\newcommand}

\nc{\beq}{\begin{equation}}  
\nc{\eeq}{\end{equation}}  
\nc{\beqa}{\begin{eqnarray}}  
\nc{\eeqa}{\end{eqnarray}}  
\nc{\bit}{\begin{itemize}}  
\nc{\eit}{\end{itemize}}  

\newcommand{\eg}{{\it e.g.}}
\newcommand{\numsig}{N_\sigma}
\newcommand{\numconsec}{N_\text{consec}}

\usepackage{floatrow}
\newfloatcommand{capbtabbox}{table}[][\FBwidth]

\usepackage{blindtext}

\title{ \bf
 Microlensing of X-ray Pulsars:  \\ 
\begin{large} 
  a Method to Detect Primordial Black Hole Dark Matter
  \end{large} }
\author{\large Yang Bai$^{a,b}$ and Nicholas Orlofsky$^{a}$}
\date{\small \it 
$^a$Department of Physics, University of Wisconsin-Madison, Madison, WI 53706, USA  \\
$^b$Theoretical Physics Department, Fermilab, Batavia, IL 60510, USA
}

\begin{document}

\maketitle

\setlength{\parskip}{0.2ex}

\begin{abstract}
Primordial black holes (PBHs) with a mass from $10^{-16}$ to $10^{-11}\,M_\odot$ may comprise 100\% of dark matter. Due to a combination of wave and finite source size effects, the traditional microlensing of stars does not probe this mass range. In this paper, we point out that X-ray pulsars with higher photon energies and smaller source sizes are good candidate sources for microlensing for this mass window. Among the existing X-ray pulsars, the Small Magellanic Cloud (SMC) X-1 source is found to be the best candidate because of its apparent brightness and long distance from Earth. We have analyzed the existing observation data of SMC X-1 by the RXTE telescope (around 10 days) and found that PBH as 100\% of dark matter is close to but not yet excluded. Future longer observation of this source by X-ray telescopes with larger effective areas such as AstroSat, Athena, Lynx, and eXTP can potentially close the last mass window where PBHs can make up all of dark matter. 
\end{abstract}

\thispagestyle{empty}  
\newpage  
  
\setcounter{page}{1}  

\begingroup
\hypersetup{linkcolor=black}
\endgroup

\newpage

\section{Introduction}\label{sec:Introduction}

Primordial black holes (PBHs) were proposed as a macroscopic dark matter (DM) candidate a few decades ago~\cite{Carr:1974nx}. They can be formed in simple inflationary models and do not require new physics below the inflationary scale (see, \eg, Refs.~\cite{Carr:2016drx,Sasaki:2018dmp} for reviews). Because of their simplicity as a DM candidate, it is necessary to search for PBHs with all possible masses. Although there are many theoretical and experimental efforts to search for PBHs, there is still a mass window from around $10^{-16}$ to $10^{-11}~M_\odot$ within which PBHs can still compose all of dark matter. It is the purpose of this paper to identify a search method to find or constrain PBHs in this mass window. 

In order to be stable on cosmological time scales and evade extragalactic gamma ray bounds from evaporation, PBHs must have mass $M \gtrsim 10^{17}~\text{g}$ or  $10^{-16}~M_\odot$~\cite{Carr:2009jm}. 
Bounds from evaporation into cosmic rays can set stronger limits for a subdominant PBH DM fraction, though these bounds are slightly weaker than gamma ray bounds when PBHs comprise all of DM \cite{Boudaud:2018hqb}.
However, ``small'' PBH masses remain relatively unconstrained for many orders of magnitude in mass above these bounds.  Previous searches for small-mass PBHs include microlensing \cite{Paczynski:1985jf} of stars in M31 using the Subaru/HSC telescope \cite{Subaru} and femtolensing of gamma ray bursts (GRBs) \cite{1992ApJ...386L...5G} using the Fermi GBM detectors \cite{Barnacka:2012bm}.  The Subaru/HSC study was limited by wave effects~\cite{1992ApJ...386L...5G} and finite source size effects~\cite{1994ApJ...430..505W}  and can only probe PBH masses $M \gtrsim 10^{22}~\text{g}$ or $10^{-11}~M_\odot$.  Regarding the study of Fermi GBM data, it was pointed out in Ref.~\cite{Katz:2018zrn} that GRBs cannot at present set bounds on PBHs because the size of the GRB gamma-ray emitting region is too large compared to the Einstein radius of the lens.
Future observations may eventually probe approximately $M \in \left[10^{17},10^{19}\right]~\text{g}$ if GRBs with small enough source size are observed.  Other potential constraints in this regime come from neutron star capture \cite{Capela:2013yf} and white dwarf destruction \cite{Graham:2015apa}, both of which face astrophysical uncertainties including the DM abundance in globular clusters \cite{Conroy:2010bs,Ibata:2012eq,Naoz:2014bqa,Popa:2015lkr}. 
Other microlensing studies at larger masses above $10^{24}~\text{g}$ include  MACHO \cite{Allsman:2000kg}, EROS \cite{Tisserand:2006zx}, OGLE \cite{Wyrzykowski:2011tr}, Kepler \cite{Griest:2013aaa}, caustic crossing \cite{Oguri:2017ock}, and quasar microlensing \cite{Mediavilla:2017bok}.
Thus, a potential window exists for PBHs to be all of DM with mass in the approximate range of $M \in \left[10^{17},10^{22}\right]~\text{g}$ or  $\left[10^{-16},10^{-11}\right]\,M_\odot$.

In this paper, we investigate whether any astrophysical object could make a suitable source to search for gravitational lensing due to PBHs in this mass window.  A few criteria for a source to serve as a good (micro-)lensing object include:
{\it i}) a large photon energy with sufficient photon counts to reduce the wave effects of lensing; {\it ii}) a small geometric size compared to the Einstein radius such that the finite source size effects are small; {\it iii}) a long distance from the telescopes around the Earth to increase the optical depth or the number of possible lensing events; {\it iv}) a large steady photon flux such that a sudden brightness magnification can be easily identified. 

For the first condition {\it i}), the wave effects becomes important when $4G_N M E_\gamma \lesssim 1$~\cite{1992ApJ...386L...5G} or $E_\gamma \lesssim 1/(4 G_N M) = 0.66\,\mbox{keV}\,\times\,\left(10^{20}\,\mbox{g}/M\right) \,$, where $G_N$ is Newton's gravitational constant and $E_\gamma$ is the lensed photon energy.
This leads us to consider sources emitting in the X-ray spectrum with energy above 1 keV, where we may ignore the wave effect for $M \gtrsim \text{few} \times 10^{20}$~g, but not for a smaller mass. In our full analysis, we will take the wave effects into account to determine the minimum mass that can be probed.  

The second condition {\it ii}) points towards using highly compact sources.  To have a rough understanding of the finite source size effects, we can compare the source size and the Einstein radius when both are projected on the lens plane. Defining $x = D_{\rm OL}/D_{\rm OS}$ as the ratio of the observer-lens angular diameter distance, $D_{\rm OL}$, over the observer-source angular diameter distance, $D_{\rm OS}$, the source radius $R_{\rm S}$ is reduced to $x R_{\rm S}$ after projection to the lens plane. The Einstein radius has~\footnote{Because we are working on galactic scales, we assume $D_{\rm OL}+D_{\rm LS}=D_{\rm OS}$, with $D_{\rm LS}$ the lens-source angular diameter distance.}
\beqa \label{eq:Einstein-radius}
r_{_{\rm E}}(x) = \sqrt{4 \, G_N\, M\,x\,(1-x)\,D_{\rm OS}} =
(107\,\mbox{km}) \times \left(\frac{\sqrt{x(1-x)}}{1/2} \right) \, \left(\frac{D_{\rm OS}}{50\,\mbox{kpc}}\right)^{1/2} \left(\frac{M}{10^{19}\,{\rm g}}\right)^{1/2}\,.
\eeqa
The ratio of the source and Einstein radii is given by
\beqa \label{eq:as}
a_{\rm S}(x) = \frac{x R_{\rm S}}{r_{_{\rm E}}(x)} \approx \left(0.1\right) \times  \left(\frac{x}{\sqrt{x(1-x)}} \right)\left(\frac{R_{\rm S}}{20\,\mbox{km}}\right)\left(\frac{50\,\mbox{kpc}}{D_{\rm OS}}\right)^{1/2} \left(\frac{10^{19}\,{\rm g}}{M}\right)^{1/2} \,,
\eeqa
which suggests a very compact source object like a neutron star or stellar mass black hole in order to have $a_{\rm S}(x) \ll 1$ for $x=\mathcal{O}(1)$.

The third and fourth conditions are somewhat at odds---a large distance puts more lenses between the source and the telescope, but it also decreases the source apparent  brightness.  Balancing these turns out to favor sources towards the outer reaches of the Milky Way  halo, \eg, in Milky Way satellite galaxies.

In the next section, we motivate why X-ray binary pulsars satisfy these conditions and determine the best candidate source pulsars.  The following three sections detail calculations of the lensing event rate and magnification, including wave and finite source size effects.  Section \ref{sec:experiments} presents current and prospective experimental bounds.  We conclude in Section \ref{sec:conclusion}.

\section{X-ray pulsars as lensing sources}\label{sec:source}

Among the X-ray sources with emitted photon energy around 1-100 keV, X-ray binaries are potential good candidates for lensing because the X-ray emission region can be relatively small. Most X-ray binaries consist of a compact stellar remnant and a nearby relatively normal donor star. Typically, the compact objects are either a neutron star ($\sim$ 1-2 $M_\odot$) or a black hole  ($\sim$ 5-15 $M_\odot$)~\cite{Zhang:2010qr,Casares:2017jah}. 
The matter from the donor gravitationally infalls into the compact object, forming an accretion disk. X-rays are emitted according to the accretion mechanism~\cite{Shakura:1972te,Rappaport:2003un}, with an X-ray emission region within a factor of few times the neutron star radius or the black hole Schwarzschild radius. For an X-ray pulsar with a solar-mass neutron star as the accretor, the hard X-rays are mainly emitted from the accretion column with a polar cap radius of $0.1\,R_{\rm NS}$ and a cylinder height of $\lesssim R_{\rm NS}$, with $R_{\rm NS}\approx 10$~km denoting the neutron star radius \cite{Hickox:2004fy}. Since the emitting direction of the hard X-rays is approximately perpendicular to the column height, the source size is anticipated to be less than around the neutron-star radius, or $R_{\rm S} \lesssim R_{\rm NS}$, and is generically below 100~km. Given the uncertainty on the current understanding of the source size, we will include the finite source size effects for $R_{\rm S}$ up to 100 km and choose a fiducial value of $R_{\rm S}=20$~km for our later analysis. The brightest X-ray black hole binaries in general are more massive and thus have a larger emitting area and more important finite source size effects. 

The observed X-ray spectrum for an X-ray pulsar is dominated by two features: direct emission from its accretion column as described above and reprocessing of column X-rays by its accretion disk~\cite{Hickox:2004fy}. The reprocessing dominates the soft energy spectrum below about 1 keV, while the accretion column dominates above about 2 keV for the pulsars in our study~\cite{Hickox:2005nd,Hung:2010cf}.  While the source size of the reprocessed X-rays is potentially large, as discussed above the accretion column is smaller.  Thus, it is important to limit any lensing search using these sources to energies greater than 2 keV, which by coincidence aligns nicely with the energy where wave effects become less important for PBH mass around $10^{20}$~g---the mass region we wish to probe.

Among all the X-ray pulsars, in order to satisfy the conditions $\it iii$) and $\it iv$) in Section~\ref{sec:Introduction}, we focus on the most distant bright sources. It is straightforward to identify the X-ray pulsars either in the Large or Small Magellanic Clouds (LMC or SMC) with a distance of $50$-$65$~kpc as the potential good sources~\cite{Casares:2017jah}. Furthermore, to have a large value of observed photon counts per second, we eventually identify SMC X-1 and LMC X-4 as the two ``good" sources to search for lensing events by PBH's and concentrate on SMC X-1 for quantitative analysis.

\section{Estimation of optical depth and averaged time interval}\label{sec:optical-depth}

Before we introduce the formulas to calculate the event rate with both wave and finite source size effects, we first estimate the optical depth and average time interval between lensing events~\cite{Paczynski:1985jf}. We will use more precise formulas in Section~\ref{sec:event-rate} for our final sensitivity study. To estimate the optical depth for PBH DM lensing a source in SMC and LMC, we use the isotropic Einasto profile~\cite{Graham:2006ae} as the dark matter density in our galaxy
\beqa
\rho_{\rm DM}(r) = \rho_{\odot} \, e^{ -\frac{2}{\beta} \left[(r/r_s)^\beta- (r_\odot/r_s)^\beta \right] } \,,
\eeqa
with $\rho_\odot = 0.4\,\mbox{GeV}/\mbox{cm}^3$, $r_s = 20$~kpc, $r_\odot = 8.5$~kpc and $\beta = 0.17$. Other dark matter profiles will only introduce a small perturbation for later results. In our analysis, we will also conservatively ignore the dark matter  contributions from SMC and LMC, which only increase the optical depth by around 10\% in the point-like source case and even smaller for the finite source size case. 

For a point-like source and ignoring wave effects, the optical depth,  or the probability for a source to be within $y_T$ Einstein radii of a foreground PBH lens, is simply
\beqa
\label{eq:optical-depth}
\tau = f_{\rm PBH}\, \int_0^{1} dx\,D_{\rm OS}\,\frac{\rho_{\rm DM}(x\,\vec{r}_{\rm S})}{M}\, \pi\,r_{\rm E}^2(x) \, y_T^2 ~~.
\eeqa
Here, $f_{\rm PBH}$ is the fraction of PBH contributions to the total DM energy density and $y_T$ is the threshold PBH distance from the source line of sight in units of $r_{_{\rm E}}$---its value depends on the required magnification factor. The integrand of (\ref{eq:optical-depth}) is independent of the lens mass, but has a quadratic dependence on the source distance [see Eq.~\eqref{eq:Einstein-radius}]. For the source SMC X-1, with $(\ell, \delta)=(300.41^\circ, -43.56^\circ)$ and at a distance of $D_{\rm OS}=d_{\rm SMC-X1}\approx 65$~kpc~\cite{Hilditch:2004pz,Keller:2006ek} from Earth, the optical depth is $8.4\times 10^{-7}$. For LMC X-4 with $(\ell, \delta)=(276.33^\circ, -32.53^\circ)$~\cite{LMC-X-4} and $D_{\rm OS}=d_{\rm LMC}=50$~kpc in distance, the optical depth is $5.5\times 10^{-7}$. The optical depths to other X-ray pulsars that are within our galaxy~\cite{x-ray-pulsars} are only a few percent of or even smaller than the optical depths for SMC and LMC sources, so we will not include them in our analysis. 

To have a rough estimate of the lensing event rate or the averaged time interval between two events, we adopt the approximate formula in Ref.~\cite{Paczynski:1985jf} 
\beqa
\langle \Delta t \rangle &=& \Gamma^{-1} \approx \frac{\pi}{2}\,\frac{t_{\rm E}}{\tau}\,f_{\rm PBH}^{-1}\,y_T\, \nonumber \\
&\approx& (11\,\mbox{days})\,\times f_{\rm PBH}^{-1}\, y_T^{-1}\, \left(\frac{\sqrt{x(1-x)}}{1/2} \right) \, \left(\frac{D_{\rm OS}}{65\,\mbox{kpc}}\right)^{1/2} \left(\frac{M}{10^{19}\,{\rm g}}\right)^{1/2} \,.
\label{eq:Deltat-opticaldepth}
\eeqa
Here, we have used $\tau=8.4\times 10^{-7}$ for SMC X-1. The Einstein radius crossing time is $t_{\rm E} \approx r_{_{\rm E}}(x=1/2)/v_\perp \approx 0.50\,\mbox{s}$ for $D_{\rm OS} = 65$~kpc, $M=10^{19}$~g and the PBH perpendicular speed around $v_\perp \approx 240$~km/s~\cite{Nesti:2013uwa}. In the situation with negligible background events, an observation of this X-ray source with a length of $\mathcal{O}(10\,\mbox{days})$ could constrain PBH as 100\% of DM. In the following section, we will include both the wave and finite source size effects and make a more realistic estimation of the event rate. 

\section{Wave optical lensing for a finite source size}\label{sec:wave}

For a source emitting primarily with X-ray energy of $\mathcal{O}(1$-$10\,\mbox{keV})$, wave effects must be taken into account in order to probe a lower PBH mass range $\lesssim 10^{19}$\,g.  For a point-like source, the magnification  factor $\mu$ is given by~\cite{Matsunaga:2006uc}
\beqa
\mu(w, y) = \frac{\pi\, w}{1 - e^{-\pi\, w}}\,\left| _1 F_1\left( \frac{i}{2}\,w, 1; \frac{i}{2}\,w \, y^2 \right) \right|^2 \,,
\eeqa
with $w \equiv 4 G_N M E_\gamma$~\footnote{For sources near or in our galaxy, we have ignored the redshift factor for the lens distance.} and $y(x)\equiv d_s(x)/r_{_{\rm E}}(x)$, with $d_s(x)$ as the tangential distance between the source and lens. Note that the mass dependence in $w$ comes from the black hole Einstein radius.  This formula is valid for any lens of mass $M$ so long as its radius is less than the Einstein radius. In the limit of $y=0$, the hypergeometric function $_1 F_1$ approaches 1 and the maximal magnification is simply the prefactor, $\mu^{\rm max} = \pi \, w/(1 - e^{-\pi w})$. For a general $y$, we can also calculate the two limits of $\mu$ in terms of $w$, which are
\beqa \label{eq:mu-two-limits}
\mu(w, y) = 
\begin{cases}
 1 + \frac{\pi\,w}{2} + \frac{w^2}{12} (\pi^2 - 3 y^2) & \mbox{for} ~~~ w \ll 1 \\
 \frac{1}{y\sqrt{4+y^2}}\left\{ 2+y^2 + 2 \sin{\left[ w \left( \frac{1}{2} y \sqrt{4+y^2} + \log{\left|\frac{\sqrt{4+y^2}+y}{\sqrt{4+y^2}-y} \right|} \right) \right]} \right\}   & \mbox{for} ~~~ w \gtrsim y^{-1} 
\end{cases}	\,.
\eeqa
So, when the wave effect is important with $w \rightarrow 0$, $\mu \rightarrow 1$ and there is no magnification. 

\begin{figure}[thb!]
\begin{center}
\includegraphics[width=0.55\textwidth]{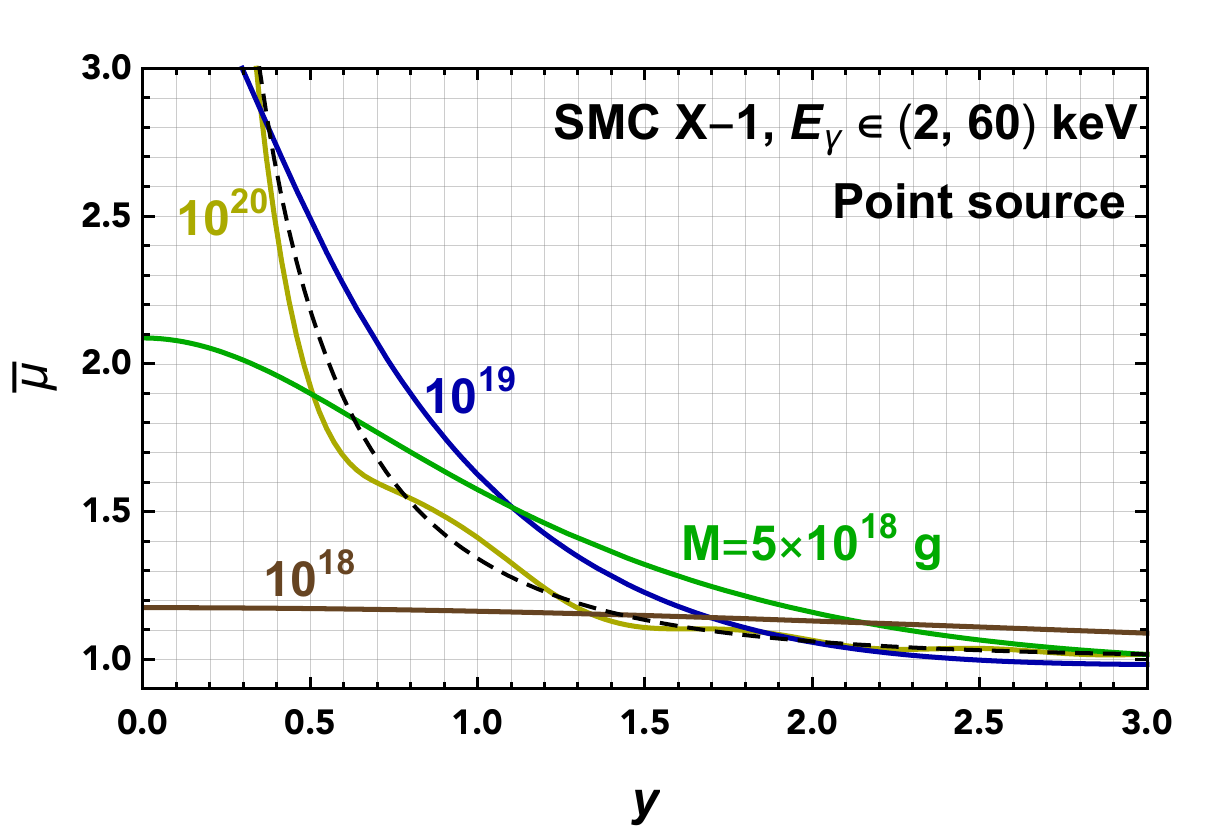} 
\caption{\label{fig:muSMC} 
The averaged magnification factor $\overline{\mu}$ for a range of photon energies as a function of $y$ defined as the ratio of the tangential source-lens separation in the lens plane over the Einstein radius. The source is assumed to be point-like for this plot. The black and dashed line is in the infinite mass limit.
}
\end{center}
\end{figure}

For a specific source, one could calculate the averaged magnification factor after integrating out a range of energy. For a source energy spectrum of $F(E_\gamma)$, we define
\beqa \label{eq:mu-energy-ave}
\overline{\mu}(y) \equiv \frac{\int^{E_{\rm max}}_{E_{\rm min}}\,d E_\gamma\,F(E_\gamma)\,\mu\left[w(E_\gamma), y\right] }{\int^{E_{\rm max}}_{E_{\rm min}}\,d E_\gamma\,F(E_\gamma)} ~~.
\eeqa
When analyzing the data for a specific telescope, one should also include the energy-dependent effective acceptance area of the telescope $A(E_\gamma)$ by making the replacement $F(E_\gamma) \rightarrow F(E_\gamma)\,A(E_\gamma)$. The hard energy spectrum for an X-ray pulsar usually follows a power-law with an exponential cutoff. For SMC X-1, we take $F(E_\gamma) = E_\gamma^{-0.93}$ for $E_\gamma \leq 6$~keV and $E_\gamma^{-0.93}\,e^{-(E_\gamma - 6~\text{keV})/7.9~\text{keV}}$ for $E_\gamma > 6$~keV~\cite{Neilsen:2004eb}. The averaged energy for the range from 2 to 60 keV is $\langle E_\gamma \rangle = 6.8$~keV.   Integrating out this energy range, we show the magnification factors for different PBH masses in Fig.~\ref{fig:muSMC}. It is clear from this figure that the magnification factor decreases as mass decreases and the wave effect becomes more important. However, this decrease is not monotonic. For instance, the corresponding value of $y$ for $\overline{\mu} = 1.8$ is the larger for $M=10^{19}$~g than for $10^{20}$~g. For $M=10^{18}$~g, the maximum magnification factor is slightly below 1.2. So, there may exist a threshold PBH mass under which lensing is undetectable. To get around this, one may consider increasing $E_{\rm min}$ to reduce the wave effect at the cost of reducing the total photon counts and increasing statistical errors. We will come back to this point when we analyze the real data.

Having discussed the wave effects, we now include the finite source size effect. Given our limited understanding of the source spatial properties, we simply assume a two-dimensional Gaussian distribution with the source size of $R_{\rm S}$ in each direction. The source intensity is $W(\vec{\chi}) \propto \mbox{exp}\left( - |\vec{\chi}|^2/2R_S^2\right)$, where $\vec{\chi}$ is the two-dimensional vector with respect to the source center. After integrating out the angular variable, one rewrites the magnification factor for a fixed energy~\cite{Goodman}
\beqa
\mu\left[ w(E_\gamma), a_{\rm S}(x), y(x) \right] = a_{\rm S}^{-2} \,e^{- y^2/(2 a_{\rm S}^2)}\, \int^\infty_0 dz\,z\,e^{-z^2/(2 a_{\rm S}^2)}\,I_0\left(y\,z/a_{\rm S}^2\right) \,\mu(w, z) \,.
\label{eq:mu-finitesource}
\eeqa
Here, the dimensionless parameter, $a_{\rm S}(x)$, is defined in Eq.~\eqref{eq:as} and proportional to the source size, $R_{\rm S}$. The function $I_0(z)$ is the zeroth-order modified Bessel function. Similarly to Eq.~\eqref{eq:mu-energy-ave}, one can also calculate the energy-averaged $\overline{\mu}$ by integrating over the relevant energy range. 

\begin{figure}[thb!]
\begin{center}
\includegraphics[width=0.47\textwidth]{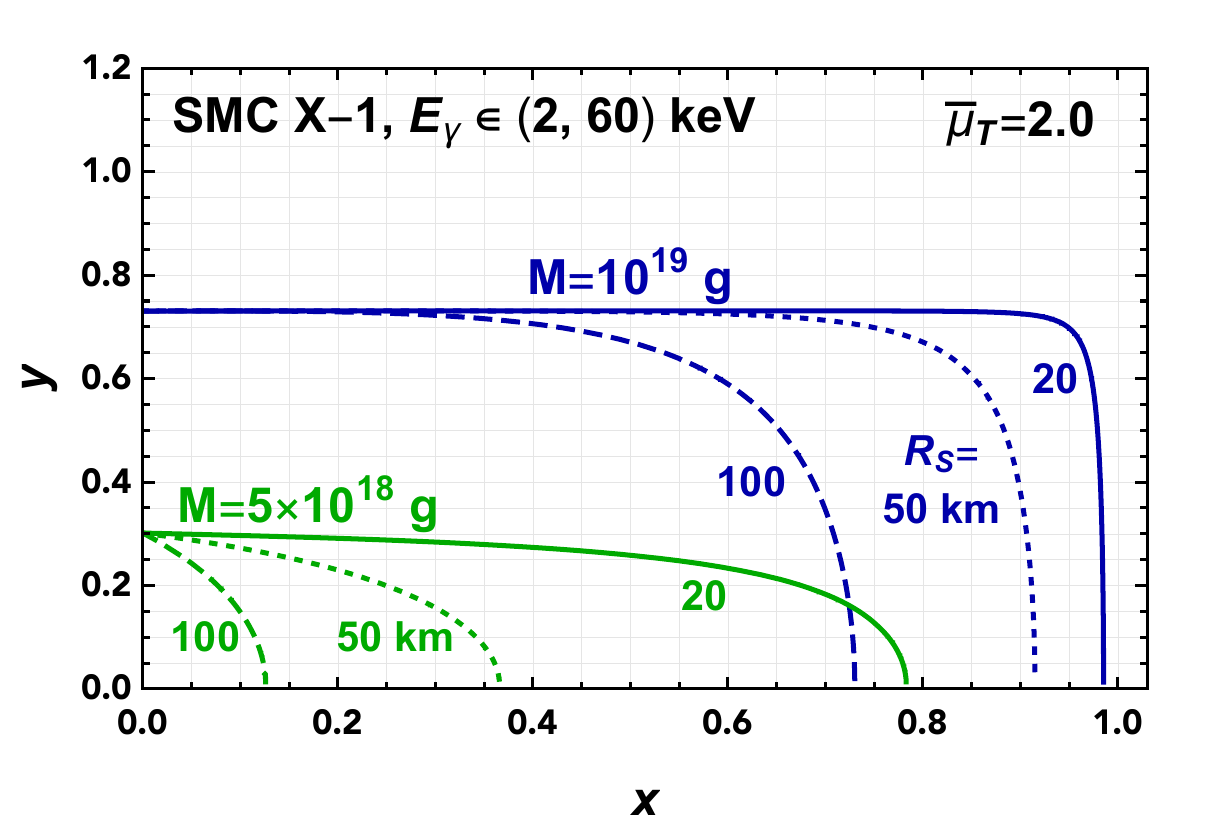} \hspace{6mm}
\includegraphics[width=0.47\textwidth]{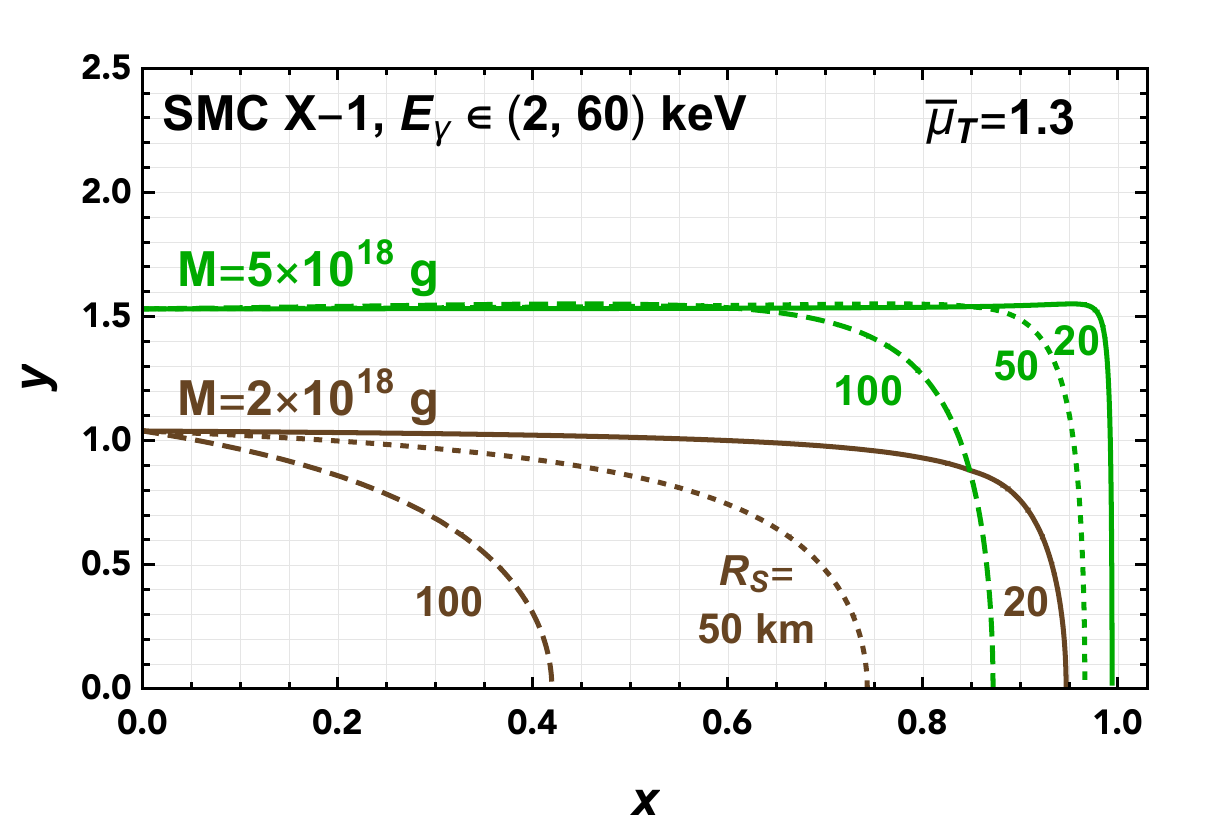}  
\end{center}
\caption{\label{fig:y-x-muT} 
The allowed parameter space (below the curves) in $x$-$y$ after including both wave and finite source size effects for two different energy-averaged magnification factors $\overline{\mu}_T=2.0$ (left panel) and 1.3 (right panel).
}  
\end{figure}

Requiring threshold values of $\overline{\mu}_T = 2.0$ or 1.3, we show the allowed parameter space in the $x$-$y$ plane in Fig.~\ref{fig:y-x-muT}. For a larger value of magnification factor (left panel), the finite source size effect is more dramatic. As the source size increases, the allowed range in $x$ shrinks, which results in a smaller optical depth and a longer observation time required to place a limit. For a small value of $x$, the finite source size effect is not important because $a_{\rm S}(x) \rightarrow 0$ as $x\rightarrow 0$. The allowed range in $y$ increases when the threshold magnification $\overline{\mu}_T$ decreases, and we have already seen from the optical depth that a larger value for $y_T$ increases the rate of lensing. So, the final sensitivity when searching for PBH microlensing events depends on the choice of $\overline{\mu}_T$ for which lensing can be distinguished from normal source fluctuations.  We will determine $\overline{\mu}_T$ based on the variance in the count rate from telescope observations.

\section{Event rate}\label{sec:event-rate}
To calculate the event rate, we take into account the dark matter velocity distribution in our galaxy. Ignoring the small effects of the source motion~\cite{1991ApJ...372L..79G}, the differential event rate is given by~\cite{1991ApJ...372L..79G,Subaru} 
\beqa
\frac{d\Gamma}{d\hat{t}} = f_{\rm PBH}\times 2\, \int^{x_{\rm max}}_0 dx\,D_{\rm OS}\,\frac{\rho_{\rm DM}(x\,\vec{r}_{\rm S})}{M}\, \int^{y_T(x)}_0 \, \frac{dy}{\sqrt{y_T(x)^2 - y^2}}\, \frac{v_r^4}{v_c^2} \,e^{- v_r^2/v_c^2 } ~~.
\eeqa
Here, $\hat{t}$ is the timescale of the microlensing event; $v_r$ is the velocity of the PBH in the lens plane and is related to $\hat{t}$ by $v_r = 2\,r_{_{\rm E}}(x) \sqrt{y_T(x)^2 - y^2}/\hat{t}$; $y_T(x)$ is the threshold source-lens distance to have $\overline{\mu} > \overline{\mu}_T$ as displayed in Fig.~\ref{fig:y-x-muT}; $x_{\rm max} \in \left[0,1\right]$ is the upper value of $x$ depending on the source size as in Fig.~\ref{fig:y-x-muT}. The velocity $v_c$ is the velocity dispersion in our galaxy, which is taken to be approximately the circular velocity. For our analysis, we simple take this velocity to be $v_c \approx 240\,\mbox{km}/\mbox{s}$ for a wide range of locations away from the center of the galaxy~\cite{Nesti:2013uwa,doi:10.1093/mnras/stw2775}.

\begin{figure}[h!]
\begin{center}
\includegraphics[width=0.48\textwidth]{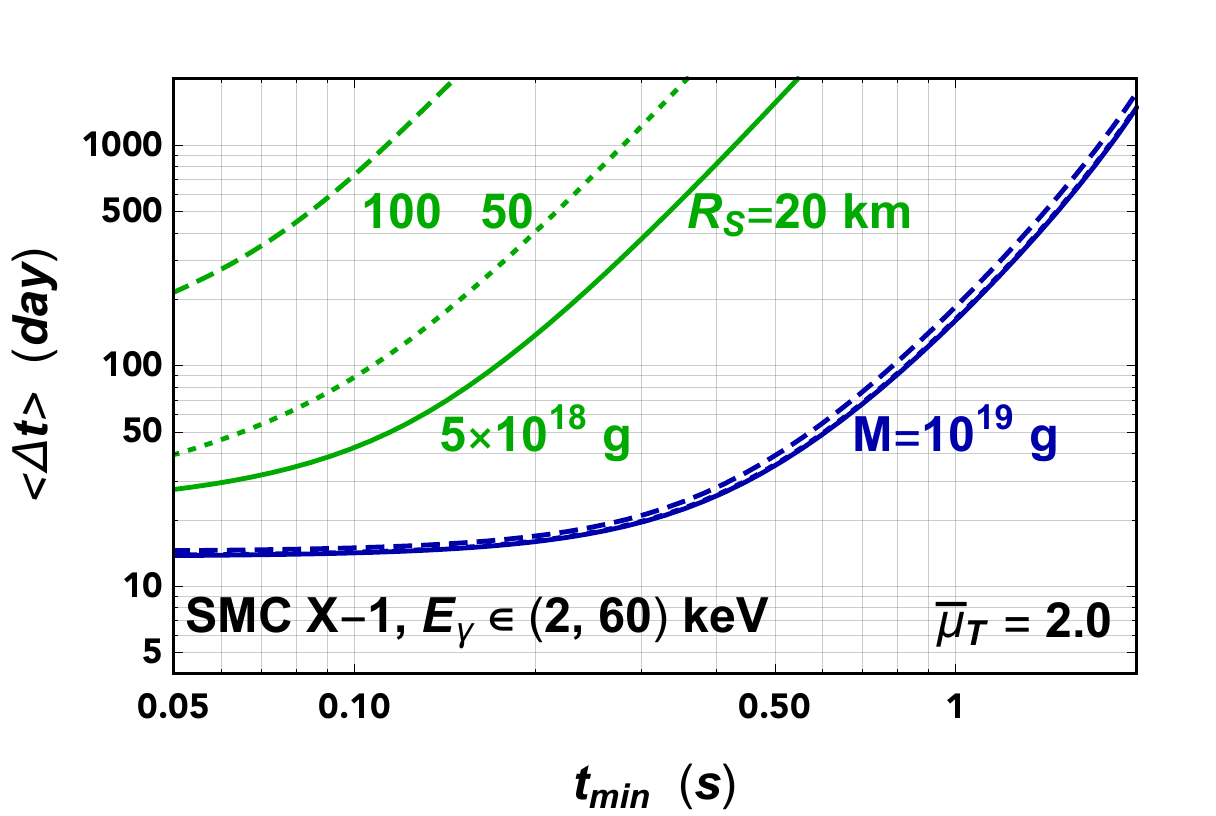} \hspace{6mm}
\includegraphics[width=0.47\textwidth]{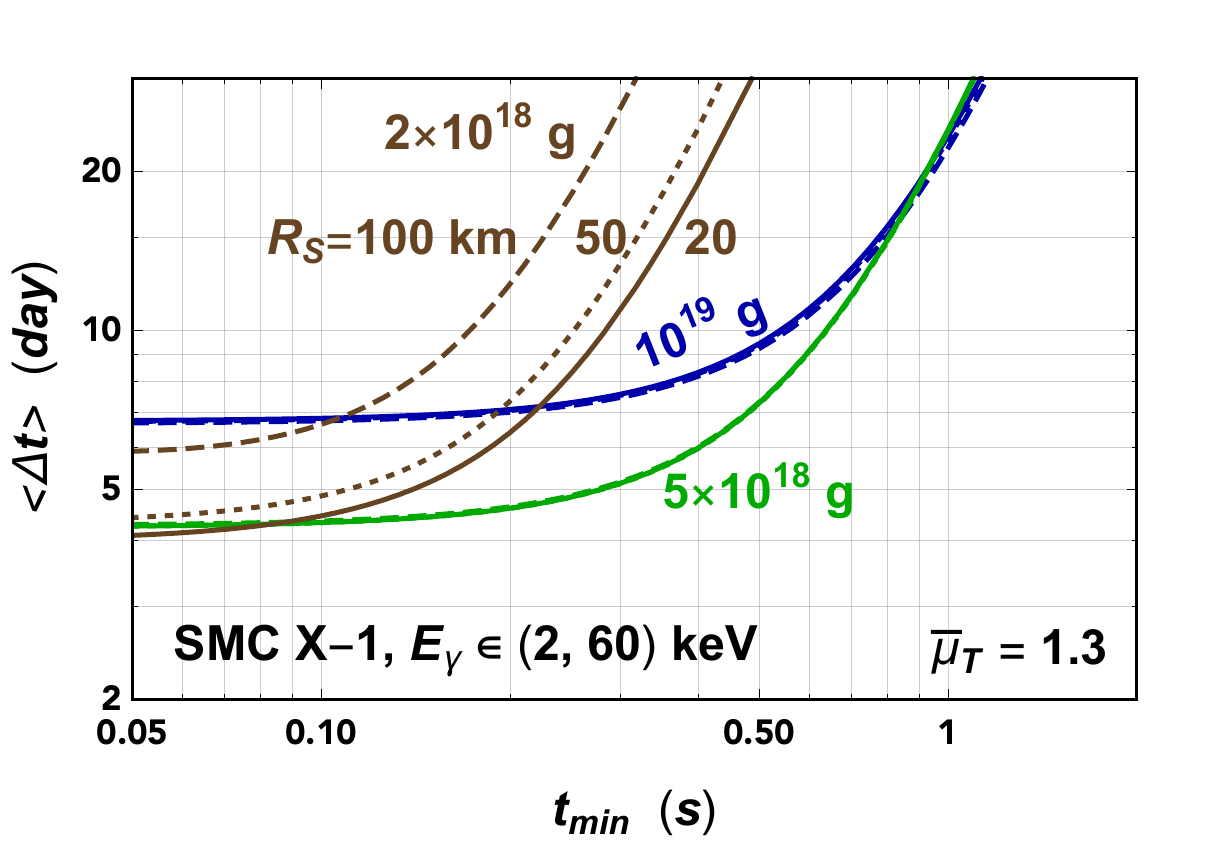}  
\end{center}
\caption{\label{fig:dGammadt} 
The averaged time interval between two lensing events as a function of the minimum of event timescale, $t_{\rm min}$. 
Various source sizes are displayed with different line textures; some of these source-size curves overlap, indicating that source size is not important for that particular $M$ and $\overline{\mu}_T$.
}  
\end{figure}

Depending on the experimental data, one could choose a minimum value for the lensing timescale, $t_{\rm min}$, which should be a factor of few times the time binning $t_{\rm bin}$ in order to have magnified counts for a few bins.  Then, the average time interval from one event to another is
\beqa
\langle \Delta t \rangle  = \left(\int_{t_{\rm min}}^\infty \frac{d\Gamma}{d\hat{t}}\right)^{-1} \,.
\label{eq:time-interval-avg}
\eeqa
We show this quantity as a function of $t_{\rm min}$ for different PBH masses and source sizes in Fig.~\ref{fig:dGammadt}. Again, for a smaller value of magnification factor, the finite source size effects are smaller for a fixed PBH mass. For $\overline{\mu}_T =1.3$ and $t_{\rm min}=0.3$~s, the averaged time interval is around 7 days for $M=10^{19}$~g and 5 days for $M=5\times 10^{18}$~g.

\section{Existing and future experimental constraints}\label{sec:experiments}

Having outlined the calculations for optical depth, magnification, and event rate, we now determine the most promising lensing sources and demonstrate how effectively X-ray telescopes can search for PBH lensing events.  
Between the X-ray pulsars identified above, SMC X-1 has a larger luminosity of around $1.5\times 10^{39}\,\mbox{erg}/\mbox{s}$~\cite{doi:10.1093/mnras/197.2.247} compared to LMC X-4's $4\times 10^{38}\,\mbox{erg}/\mbox{s}$~\cite{1991ApJ...381..101L}.  In addition, it is at a greater distance, giving it a larger optical depth for lensing.  It also has longer total archival observations by recent X-ray telescopes.  We thus focus on it for the remainder of this section.

Given the SMC X-1 source flux of $\sim 0.1\,\mbox{cts}/\mbox{s}/\mbox{cm}^2$ for X-ray energies above a few keV, a telescope with an effective area of $\mathcal{O}(10^4\,\mbox{cm}^2)$ is necessary to have $\mathcal{O}(100)$ counts for the time bin size of 0.1~s to constrain the magnification factor.  We now discuss telescopes that fit this criterion.

\subsection{Existing data from RXTE}\label{sec:existing}

Among the previous and current X-ray telescopes,  the Rossi X-ray Timing Explorer Proportional Counter Array (RXTE PCA) and AstroSat~\cite{doi:10.1117/12.2062667} have large enough effective areas in the energy range of interest (a total collecting area of 6500 cm$^2$). Their effective areas are approximately flat for energies above around 4 keV and drop quickly near 2 keV and above 10 keV (80 keV) for RXTE~\cite{RXTE-effective-area} (AstroSat). In our data analysis, we will take the energy-dependent area into account when we calculate the magnification factor using \eqref{eq:mu-energy-ave}. RXTE PCA also has more pointed exposure for SMC X-1, 12.65 days, than any other modern X-ray telescope. This exposure time is in the ballpark of the averaged time interval for the lensing events, shown in Fig.~\ref{fig:dGammadt}, which makes the observation of RXTE PCA on SMC X-1 very interesting to search for PBH dark matter.  

\begin{figure}
\begin{center}
\includegraphics[width=0.65\textwidth]{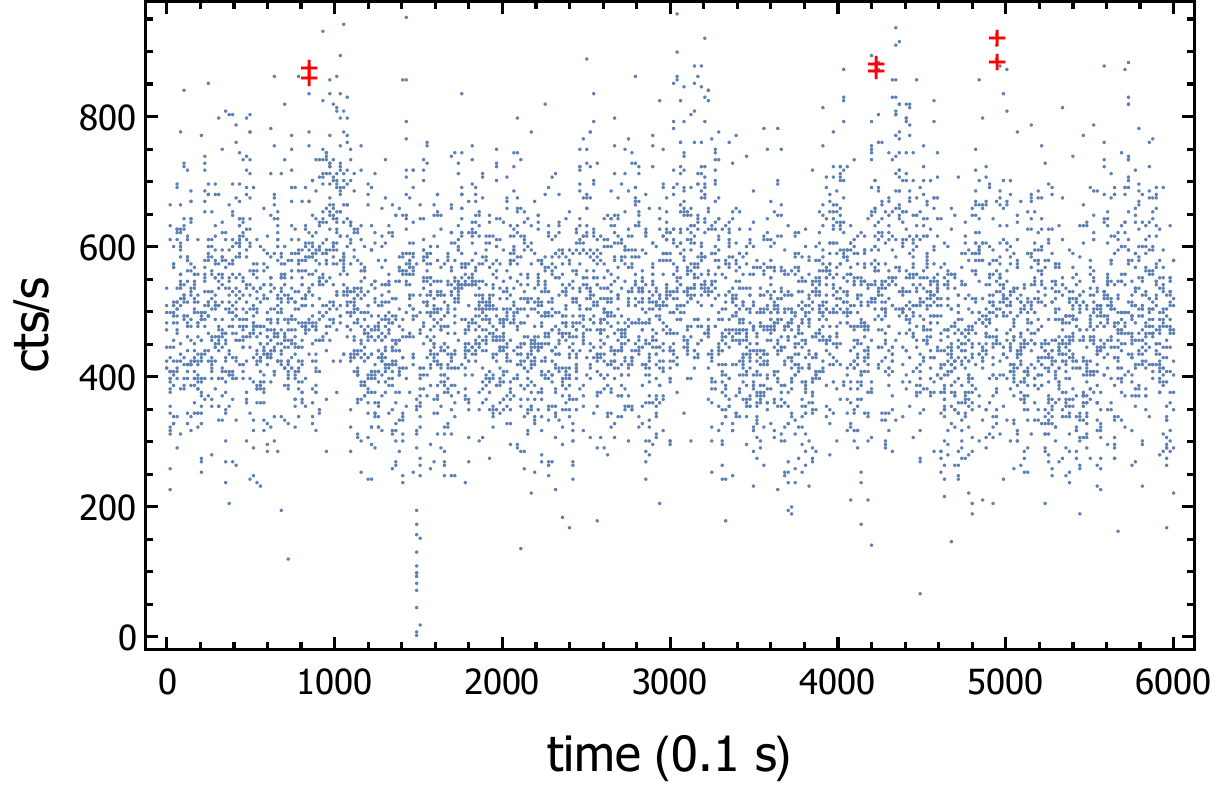} 
\end{center}
\caption{\label{fig:example} 
Example lightcurve for a 600-second portion of one observation period (observation ID P10139 with all five PCUs added) binned in 0.1 s intervals after removing the intrinsic pulsation frequency.  Red crosses indicate where there are two consecutive events exceeding the mean by at least $3\sigma$. There is no three consecutive events with $3\sigma$ deviation, which is close to the requirement for a lensing event.}
\end{figure}

We use the RXTE-specific tools in {\tt HEASOFT 6.25}~\cite{RXTE_Heasoft,Heasoft}~\footnote{$\mathtt{SEEXTRCT}$ was used to extract events with Earth elevation angle greater than 10 degrees, pointing offset less than 0.02 degrees, and time since South Atlantic Anomaly greater than 10 minutes.  All active Proportional Counter Units (PCUs) were included.  Background was estimated using $\mathtt{RUNPCABACKEST}$ with the provided bright source background model, and the background lightcurve was subtracted from the observed lightcurve.  Barycenter correction was performed with $\mathtt{FAXBARY}$.} to analyze the RXTE PCA data from the GoodXenon1 and GoodXenon2 modes. Since SMC X-1 has an intrinsic pulsation period of about 0.7\,s, we apply a Fourier transformation for the extracted lightcurves and remove the peaks associated with the intrinsic frequencies. We then perform an inverse Fourier transformation to convert the data back to obtain the pulsation-free lightcurves. The resulting lightcurves are used to estimate the apparent brightness and variability of the persistent emission (no flares or eclipses).  As an example, we show a portion of one observation period (observation ID P10139) in Fig.~\ref{fig:example}. For this observation with binning time $t_\text{bin}=0.1\,\text{s}$, the fiducial brightness is calculated to be $\mbox{B}_\text{fid}=496~\text{cts/s}$ with the standard deviation $\sigma_\text{B,fid}=123~\text{cts/s}$ in the persistent emission.  Note that $\sigma_\text{B,fid} t_\text{bin} \simeq 12 > \sqrt{\mbox{B}_\text{fid} t_\text{bin}} \simeq 7$, indicating that there is a bit more intrinsic variation than just the source's Poisson noise and pulsation period.  The additional variations likely come from the accretion mechanism or observational noise.  If this additional variation includes correlations between flux bins, this may weaken our results.

Using this information, we create selection criteria to search for rare lensing events while keeping minimal statistical background.  We may look for some number of consecutive points $\numconsec$ on the lightcurve for which the count rate is greater than some number of standard deviations $\numsig$ above the mean~\cite{Griest:2011av}.  The number of standard deviations $\numsig$ is chosen such that the number of expected statistical background occurrences is much smaller than unity for the entire observation period. In other words, the probability for a particular set of $\numconsec$ consecutive bins to be all above a given threshold $\numsig$ ought to obey
\beq
p \ll \frac{t_\text{bin}}{t_\text{obs}} = 1.16 \times 10^{-7} \times  \left(\frac{10~\text{days}}{t_\text{obs}}\right) \left(\frac{t_\text{bin}}{0.1~\text{s}}\right) ~,
\label{eq:pvalue}
\eeq
where $t_\text{obs}$ is the total observation time. Note that $p$ refers to a {\it particular} set of $\numconsec$ bins, as opposed to any set of $\numconsec$ bins in the entire dataset.  For example, assuming a Gaussian distribution and all points uncorrelated, $p=[1-\Phi(\numsig)]^{\numconsec}$ where $\Phi$ is the cumulative probability function for a Gaussian with mean zero and standard deviation one; then,
the probability to have three consecutive time bins with over $3\sigma$ fluctuation is $p=2.5 \times 10^{-9}$, while the probability for having two consecutive time bins with $4\sigma$ is $p=1.0\times 10^{-9}$. If the points have additional correlations in time, one could either impose a more stringent statistical requirement or diagnose the to-be-found ``interesting" events closer by examining their light curves and energy spectra. In practice, we require $\numsig$ just large enough to saturate a factor of $1/20$ times the number in the right side of Eq.~\eqref{eq:pvalue}. The actual bounds are not sensitive to the choice of factor. 

With the requirement on $\numsig$, we can then calculate the required energy-averaged magnification factor, $\overline{\mu}_T$. Say we are interested in the magnification of a particular time bin that is $n_\sigma$ standard deviations from the mean (positive or negative).  We may also wish to vary the binning time $t_\text{bin}$ and apparent source brightness ${\rm B}$. Then, the required $\overline{\mu}_T$ is 
\begin{equation}
\overline{\mu}_T = \left(1 \,+ \,\numsig \,\times\, \frac{\sigma_{{\rm B},\text{fid}}/{\rm B}_\text{fid} }{\sqrt{({\rm B}/{\rm B}_\text{fid}) (t_\text{bin}/t_\text{bin,fid})}}\right) \bigg/ \left(1 \,+ \,n_\sigma \,\times\, \frac{\sigma_{{\rm B},\text{fid}}/{\rm B}_\text{fid} }{\sqrt{({\rm B}/{\rm B}_\text{fid}) (t_\text{bin}/t_\text{bin,fid})}}\right) ~~,
\end{equation}
where ${\rm B}_\text{fid}$ is the fiducial apparent source brightness in cts/s with $\sigma_{{\rm B},\text{fid}}$ its standard deviation for a fiducial value of the binning time $t_\text{bin,fid}$. For the fiducial values given above for the RXTE PCA data, taking $\numsig=3$ and $n_\sigma=-1$, one has  $\overline{\mu}_T=2.32$ for $t_\text{bin}=0.1~\text{s}$.  Note that the magnification is required to exceed $\overline{\mu}_T$ for a time period of at least $\numconsec t_\text{bin}$.  Thus, the maximum magnification is generally greater than $\overline{\mu}_T$.  Note that this assumes each flux bin is uncorrelated.  We discuss this further in Section \ref{sec:conclusion}.

With $\overline{\mu}_T$ determined, $y_T(x)$ can be computed using Eqs.~(\ref{eq:mu-energy-ave}) and (\ref{eq:mu-finitesource}).  Then, the lensing event rate can be computed from Eq.~(\ref{eq:time-interval-avg}), which must be multiplied by the probability for there to be $\numconsec$ consecutive bins above $n_\sigma$ from the mean in the underlying source signal (before lensing effects); this probability is estimated from the data distribution.\footnote{The light curve data is nearly Gaussian; it is slightly skewed right.}  Having tried a few options, we take $\numconsec=3$ and $n_\sigma=-1$ as fixed, which tends to yield slightly better results than other possibilities. For each mass, the optimal value of $t_\text{bin}$ is determined to maximize the lensing event rate. If no lensing candidates are found and background is assumed to be nearly zero, masses and PBH abundances for which the expected number of lensing events is $\geq 3$ can be excluded at 95\% CL.

As we will now demonstrate, the present RXTE data is not sufficient to constrain $f_{\rm PBH} \leq 1$.  Therefore, we will not perform a full analysis of the RXTE data because the resulting bounds would not constrain any interesting portions of parameter space.  Rather, we give an estimate using some simplifying assumptions for what the RXTE data can exclude.  Then, in the next section, we will show how future telescopes can probe heretofore untested PBH masses.

\begin{figure}[t]
\begin{center}
\includegraphics[width=0.7\textwidth]{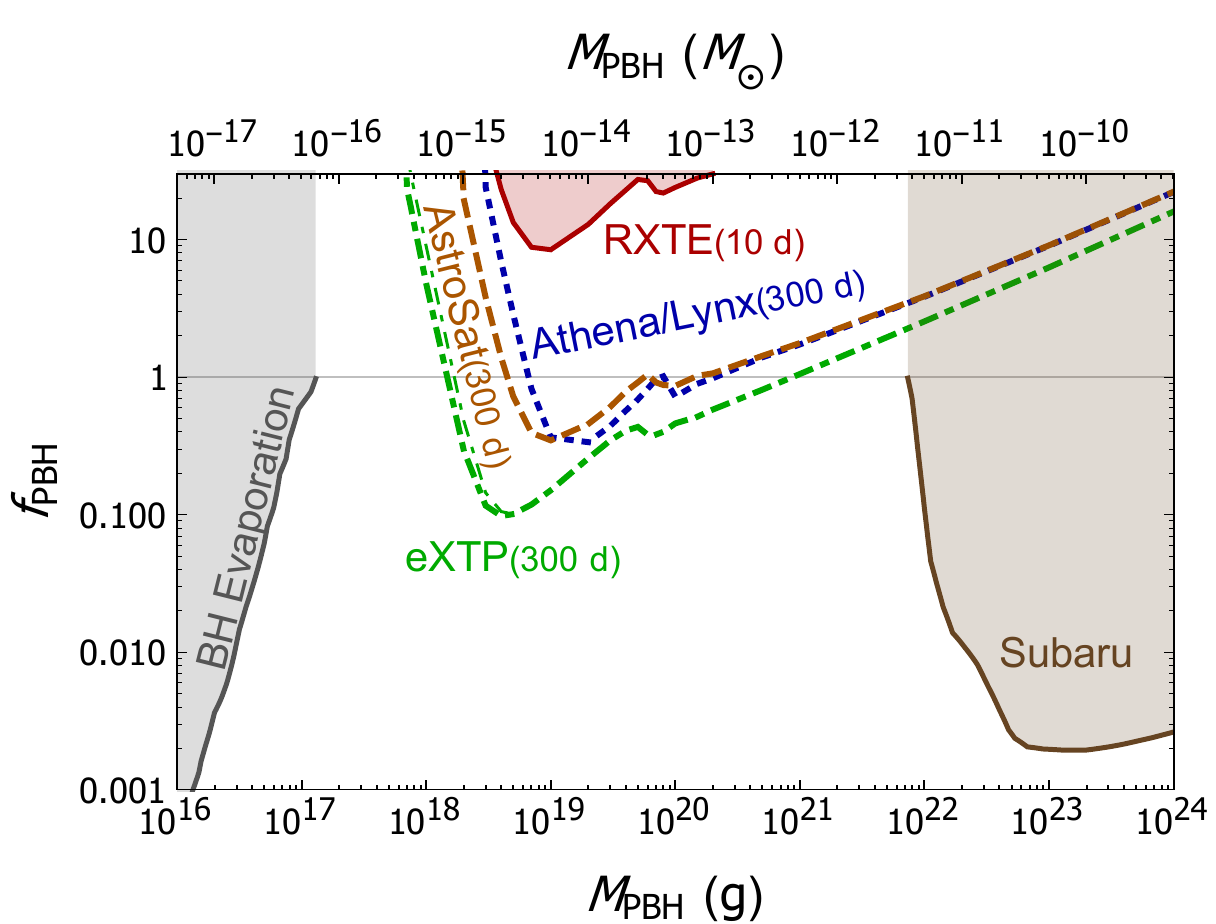} 
\end{center}
\caption{\label{fig:bounds} 
Constraints on the PBH dark matter fraction $f_\text{PBH}$ and mass $M$ at 95\% CL using around 10 days observation of SMC X-1 by RXTE. Also shown are the projected limits from future observations of SMC X-1 by AstroSat, Athena and Lynx, and eXTP. The SMC X-1 source size with the X-ray energy above 2 keV is fixed to be 20 km for all thicker curves (the thinner green dot-dashed line for eXTP has a source size of 100 km for illustration purposes). 
Finite source size effects are unimportant due to the optimization of $t_\text{bin}$ (see text for details).  
The flux measurements are optimistically assumed to be uncorrelated during persistent emission.
Also shown are extragalactic gamma ray bounds from BH evaporation \cite{Carr:2009jm} and Subaru/HSC microlensing bounds \cite{Subaru} (truncated at $f_\text{PBH}=1$).
}
\end{figure}

For the total RXTE PCA 12.65-day exposure of SMC X-1, there are about 10 days of persistent emission. We take a constant persistent count rate of $\text{B}=170~\text{cts/s/pcu}$~\cite{Raichur:2009ej}, though in reality the persistent emission varies with the superorbital period; \eg, the count rate in Fig.~\ref{fig:example} is a bit lower, while even higher rates have been observed~\cite{Rai:2018vkw}.  Future observations should focus on high points in the superorbital period to obtain the best lensing sensitivity.  We make a further simplified assumption that all 5 PCUs are active for these observations, although for many observations some of PCUs are not available.  All of these assumptions are a bit optimistic compared to actual RXTE data, but they are more realistic for future observations, which are the main focus of this work.  Since there is no microlensing-like event observed from our data analysis, we therefore set 95\% CL constraints on the PBH parameter space in $f_{\rm PBH}$ and $M_{\rm PBH}$ in the red shaded region of Fig.~\ref{fig:bounds}. We also show the gamma-ray constraints from PBH evaporation in the gray shaded region~\cite{Carr:2009jm} and the Subaru/HSC constraints from microlensing of stars in M31 in the brown shaded region~\cite{Subaru}. 

For $M_{\rm PBH} = 10^{19}~\mbox{g}$, RXTE has the most stringent constraint of $f_{\rm PBH} \lesssim 8.4$, which requires three consecutive $3.7\sigma$ time bins with $t_\text{bin}=0.08~\text{s}$ and has $\overline{\mu}_T=2.2$. For small masses, the wave effect limits the maximum attainable magnification (see Fig.~\ref{fig:muSMC}), and so the optimization procedure prefers to increase $t_{\rm bin}$ and reduce  $\overline{\mu}_T$. 
Around the threshold mass of $\sim 2\times 10^{18}$~g, the finite source size effects would become important if $\overline{\mu}_T$ were fixed (see Fig.~\ref{fig:dGammadt}).  However, a smaller $\overline{\mu}_T$ from the optimization means that finite source size effects are reduced.
On the other hand, for larger masses, the event passing time is long, which also leads to a smaller preferred  $\overline{\mu}_T$. So, $\overline{\mu}_T$ as a function of mass has a peak value located around $10^{19}~\mbox{g}$. 
The increased sensitivity around $M = 5 \times 10^{19}~\text{g}$ is the result of wave effects giving a relatively flat $\overline{\mu}(y)$ near the optimal $\overline{\mu}_T$, allowing $y_T$ to be larger near this particular mass.  The precise location of this dip depends on the source energy spectrum and the range of energies that are integrated.

Regarding the lower energy cutoff of $E_\text{min}=2~\text{keV}$, we have tested and found that the exact choice for this value has small effects on the bounds. This is because any gain from removing the influence of wave effects at lower energy is offset by a loss in apparent source brightness, which goes as $E^{-0.93}$ before including the effective area dependence.

One may consider other existing X-ray telescope data in addition to RXTE.  For example, Chandra, XMM-Newton, Suzaku, and NuSTAR all have about 2~\text{to}~5~\text{days} of SMC X-1 exposure \cite{Heasarc_browse}.  Unfortunately, their combined exposure time is less than RXTE alone.  More importantly to the present discussion, the effective area of these telescopes is smaller than RXTE when multiple PCUs are on.  With smaller count rates, a higher $\overline{\mu}_T$ and thus smaller $y_T(x)$ is necessary to pick out a lensing signal from the background fluctuations, reducing the rate of detectable lensing events.

\subsection{Projected sensitivity: AstroSat, Athena, Lynx, and eXTP}\label{sec:future}

Although the existing data from RXTE is not sufficient to constrain PBHs as 100\% of dark matter, a longer observation of this X-ray pulsar source will probe this interesting PBH mass range. RXTE ceased its science operations in 2012, but the ongoing satellite telescope AstroSat, launched in 2015, has a similar effective area to RXTE.
The orange dashed line of Fig.~\ref{fig:bounds} shows the projected 95\% confidence level (CL) limits for a 300-day observation of SMC X-1 by AstroSat, which will constrain a wide range of PHB masses from $\text{few} \times 10^{18}$\,g to $10^{20}$\,g.  Notably, only $\mathcal{O}(100~\text{days})$ of AstroSat exposure to persistent emission are necessary to begin to constrain $f_\text{PBH} < 1$.

Among the future X-ray telescopes, Athena~\cite{Athena}, Lynx~\cite{LynxTeam:2018usc}, and eXTP~\cite{Zhang:2016ach} have larger effective areas than RXTE PCA for the interesting energy range of $2$-$10$~keV for the source SMC X-1. Both Athena and Lynx have an effective area as large as 2 m$^2$ at 1 keV, while eXTP has an even larger effective area of 3.4 m$^2$ peaked between 6 and 10 keV. Taking into account the energy-dependent effective area, we show the projected limits for 300-day observations for both Athena and Lynx in the blue dotted line in Fig.~\ref{fig:bounds}, which is similar to the limits from AstroSat but with the best-sensitive point at a slightly higher mass. This is because Athena and Lynx have their effective areas peaked at a lower value of energy, which cannot probe quite as low PBH masses due to wave effects. Finally, we show the ultimate sensitivity with a 300-day exposure for an X-ray telescope with a larger effective area like eXTP in the green dot-dashed line of Fig.~\ref{fig:bounds}.  It is interesting to note that a year-long observation of the SMC X-1 will almost cover the currently unconstrained mass gap for PBH's as an explanation for 100\% dark matter. An even larger telescope like LOFT \cite{Feroci:2011jc, Vacchi:2018mnt} (a previous version of eXTP) could probe an even larger range of masses.

All of these bounds exhibit a bump in sensitivity similar to RXTE's a bit below $10^{20}~\text{g}$ due to wave effects.  The limits for large masses are set by the increasing passing time, which results in a smaller rate that scales roughly as $\Gamma \propto M^{-1/2}$ [see Eq.~\eqref{eq:Deltat-opticaldepth}].  This effect is slightly offset by increasing $t_\text{bin}$, allowing a smaller $\overline{\mu}_T$ as mass increases.

For all the bounds in Fig.~\ref{fig:bounds}, the finite source size effect is not important, contributing less than a few percent correction for $R_{\rm S}=20$~km.  The reason is that the value of $t_\text{bin}$ has been chosen at each point to maximize the sensitivity.  As a result of this, the smallest masses where bounds are possible tend to prefer larger $t_\text{bin}$ and smaller $\overline{\mu}_T$.  Smaller $\overline{\mu}_T$ delays the wave effect and finite source size effect from becoming relevant (see Figs.~\ref{fig:y-x-muT} and \ref{fig:dGammadt}), which can overcome the decrease in $\Gamma$ as $t_\text{min} \propto t_\text{bin}$ increases.  Because this optimization can be specified {\it a priori}, there is no trial factor.

\section{Discussion and conclusions}
\label{sec:conclusion}

We note that the selection criteria as presented will pick out both gravitational lens events as well as source flares.  The two can be easily distinguished.  First, in the lightcurve, flares exhibit a sharp rise followed by an exponential decay, whereas lensing events are symmetric and have a distinct shape (that varies depending on how far one is into the wave-regime).  Furthermore, the effects of each on the energy spectrum differs, and they can be distinguished by, \eg, the hardness ratio.  For larger masses with $w \gg 1$, we are in the regime of microlensing where the magnification is uniform across all energies.  In the other case, we are in the femtolensing regime~\cite{1992ApJ...386L...5G}, and the calculable energy-dependent magnification will manifest in the measured spectrum.

Aside from their pulse periods, X-ray binaries exhibit other periodic fluctuations.  In the case of SMC X-1, it has an orbital period of 3.89 days.  During part of this period, its emissions are eclipsed by its accretion disk.  Further, it exhibits a superorbital variation with period varying in the range 40 to 65 days~\cite{Hu:2013wza,Trowbridge:2007kj}, during which it oscillates between high- and low-state emission.  To maximize lensing bounds, future observations should focus on uneclipsed high-state emission.

The effects in the above two paragraphs describe many of the intrinsic source variabilities.  However, even on top of these there are more sources of variability beyond ordinary Poisson statistics, as is evident by noting that in our fiducial values, $\sigma_\text{B,fid} t_\text{bin,fid} \simeq 12$ is a bit larger than $\sqrt{\text{B}_\text{fid} t_\text{bin,fid}} \simeq 7$.  Importantly, correlations between consecutive time bins could lead to false positives in lensing searches, which could weaken the projections presented herein.  As the purpose of this study is to identify X-ray pulsars as suitable lensing targets and provide estimated projections of their lensing sensitivity, we do not attempt a full accounting of all of the mechanisms in the accretion process that may account for this.  We leave this as a topic of future study.

While we have chosen SMC X-1 as one of the most promising (and at present, most observed) cases, other X-ray binaries could contribute to future lensing bounds.  We have already mentioned LMC X-4 as another promising X-ray pulsar which has similar distance as SMC X-1, although it is a bit fainter.  Other closer X-ray pulsars within the Milky Way disk could add further to lensing bounds, although the optical depth for lensing these sources is smaller.  Finally, X-ray black holes could provide another avenue for setting bounds.  The brightest and thus most promising X-ray black holes tend to be a bit heavier than X-ray pulsars (since pulsars are limited in mass by the requirement that they not gravitationally collapse).  While these heavier black hole radii may be on the same order as the neutron star radii, the accretion and X-ray emission region may be larger owing to their larger mass.  In addition, reprocessing dominates the black hole spectra to higher energies than for the pulsars~\cite{Nowak:2000kf}.  As a result, a larger value for $E_\text{min}$ is necessary, which reduces the overall count rate.  Even before this cut, the LMC and SMC black hole binaries are dimmer than SMC X-1. Nonetheless, they may prove especially useful for limiting larger-mass lenses where finite source size effects are unimportant.  Better understanding and modeling of the source size and shape could improve the analysis in this paper.

It may also prove advantageous to have multiple X-ray telescopes observing the same source simultaneously.  This can help to distinguish non-Gaussian noise that is not associated with the intrinsic source variability---for example, cosmic rays mimicking X-rays.  It also potentially allows for parallax detection, giving another handle on distinguishing intrinsic source variation from mircolensing signals \cite{Refsdal:1993kf,Gould:1992yv,Gould:1993yv}.  If a lensing event is observed, parallax information could allow a determination of distances to the lens and the source.  Finally, it would allow the confirmation of a microlensing measurement across more than one observatory.

Another approach to set bounds at masses nearer to the edge of the Subaru/HSC bounds is to employ sources emitting in energies between X-ray and visible, namely in the ultraviolet (UV).  For example, UV stars in the M31 could be considered.  However, finite source size effects must be taken into account for UV-emitters of stellar size.  Indeed, finite source size effects were an important limiting factor in the Subaru/HSC study. A more detailed analysis may be worth pursuing. One could also consider other UV sources like type Ia supernovae~\cite{Foley:2016obj} and hot white dwarfs~\cite{white}: the former also suffers the finite source size effect, and the latter can only be observed in our Milky Way galaxy and does not have enough optical depth. 

In this paper, we have explored the potential for X-ray telescope observations of X-ray binary pulsars to probe lensing due to PBH DM with mass $M \in \left[10^{17},10^{22}\right]~\text{g}$ or  $\left[10^{-16},10^{-11}\right]\,M_\odot$, between present   BH evaporation and Subura-HSC bounds. We have identified SMC X-1 as one of the most promising candidate sources, which strikes a balance between a distant source with large optical depth and a bright source with good counting statistics.  While present data are just shy of excluding PBH in this window, adding just $\mathcal{O}(100~\text{days})$ of exposure to persistent emission by the presently-operating AstroSat telescope to the existing RXTE data can already start to probe presently unbounded PBH masses.  A future telescope with larger effective area like eXTP could probe nearly all of the open mass range with about one year of exposure. The microlensing study for PBHs in this paper can be also applied to other macroscopic dark matter candidates like dark quark nuggets~\cite{Bai:2018dxf} or axion miniclusters or stars~\cite{Kolb:1993zz}, provided they have small enough radii.

\subsubsection*{Acknowledgements}
We thank Andrey Katz and Andrew Long for discussion.  
The work is supported by the U. S. Department of Energy under the contract DE-SC0017647 and URA Visiting Scholars Program. This work was performed at the Aspen Center for Physics, which is supported by National Science Foundation grant PHY-1066293. YB also thanks the hospitality of the particle theory group of the University of Chicago.

\bibliographystyle{JHEP}
\bibliography{lense_xpulsar}

\providecommand{\href}[2]{#2}\begingroup\raggedright\begin{thebibliography}{10}

\bibitem{Carr:1974nx}
B.~J. Carr and S.~W. Hawking, {\it {Black holes in the early Universe}},  {\em
  Mon. Not. Roy. Astron. Soc.} {\bf 168} (1974) 399--415.

\bibitem{Carr:2016drx}
B.~Carr, F.~Kuhnel, and M.~Sandstad, {\it {Primordial Black Holes as Dark
  Matter}},  {\em Phys. Rev.} {\bf D94} (2016), no.~8 083504,
  [\href{http://arxiv.org/abs/1607.06077}{{\tt arXiv:1607.06077}}].

\bibitem{Sasaki:2018dmp}
M.~Sasaki, T.~Suyama, T.~Tanaka, and S.~Yokoyama, {\it {Primordial black
  holes-perspectives in gravitational wave astronomy}},  {\em Class. Quant.
  Grav.} {\bf 35} (2018), no.~6 063001,
  [\href{http://arxiv.org/abs/1801.05235}{{\tt arXiv:1801.05235}}].

\bibitem{Carr:2009jm}
B.~J. Carr, K.~Kohri, Y.~Sendouda, and J.~Yokoyama, {\it {New cosmological
  constraints on primordial black holes}},  {\em Phys. Rev.} {\bf D81} (2010)
  104019, [\href{http://arxiv.org/abs/0912.5297}{{\tt arXiv:0912.5297}}].

\bibitem{Boudaud:2018hqb}
M.~Boudaud and M.~Cirelli, {\it {Voyager-1 $e^\pm$ further constrain Primordial
  Black Holes as Dark Matter}},  \href{http://arxiv.org/abs/1807.03075}{{\tt
  arXiv:1807.03075}}.

\bibitem{Paczynski:1985jf}
B.~Paczynski, {\it {Gravitational microlensing by the galactic halo}},  {\em
  Astrophys. J.} {\bf 304} (1986) 1--5.

\bibitem{Subaru}
H.~Niikura, M.~Takada, N.~Yasuda, R.~H. Lupton, T.~Sumi, S.~More, A.~More,
  M.~Oguri, and M.~Chiba, {\it {Microlensing constraints on primordial black
  holes with the Subaru/HSC Andromeda observation}},
  \href{http://arxiv.org/abs/1701.02151}{{\tt arXiv:1701.02151}}.

\bibitem{1992ApJ...386L...5G}
A.~{Gould}, {\it {Femtolensing of gamma-ray bursters}},  {\em Astrophysical
  Journal Letters} {\bf 386} (Feb., 1992) L5--L7.

\bibitem{Barnacka:2012bm}
A.~Barnacka, J.~F. Glicenstein, and R.~Moderski, {\it {New constraints on
  primordial black holes abundance from femtolensing of gamma-ray bursts}},
  {\em Phys. Rev.} {\bf D86} (2012) 043001,
  [\href{http://arxiv.org/abs/1204.2056}{{\tt arXiv:1204.2056}}].

\bibitem{1994ApJ...430..505W}
H.~J. {Witt} and S.~{Mao}, {\it {Can lensed stars be regarded as pointlike for
  microlensing by MACHOs?}},  {\em Astrophys. J.} {\bf 430} (Aug., 1994)
  505--510.

\bibitem{Katz:2018zrn}
A.~Katz, J.~Kopp, S.~Sibiryakov, and W.~Xue, {\it {Femtolensing by Dark Matter
  Revisited}},  \href{http://arxiv.org/abs/1807.11495}{{\tt arXiv:1807.11495}}.

\bibitem{Capela:2013yf}
F.~Capela, M.~Pshirkov, and P.~Tinyakov, {\it {Constraints on primordial black
  holes as dark matter candidates from capture by neutron stars}},  {\em Phys.
  Rev.} {\bf D87} (2013), no.~12 123524,
  [\href{http://arxiv.org/abs/1301.4984}{{\tt arXiv:1301.4984}}].

\bibitem{Graham:2015apa}
P.~W. Graham, S.~Rajendran, and J.~Varela, {\it {Dark Matter Triggers of
  Supernovae}},  {\em Phys. Rev.} {\bf D92} (2015), no.~6 063007,
  [\href{http://arxiv.org/abs/1505.04444}{{\tt arXiv:1505.04444}}].

\bibitem{Conroy:2010bs}
C.~Conroy, A.~Loeb, and D.~Spergel, {\it {Evidence Against Dark Matter Halos
  Surrounding the Globular Clusters MGC1 and NGC 2419}},  {\em Astrophys. J.}
  {\bf 741} (2011) 72, [\href{http://arxiv.org/abs/1010.5783}{{\tt
  arXiv:1010.5783}}].

\bibitem{Ibata:2012eq}
R.~Ibata, C.~Nipoti, A.~Sollima, M.~Bellazzini, S.~Chapman, and E.~Dalessandro,
  {\it {Do globular clusters possess Dark Matter halos? A case study in NGC
  2419}},  {\em Mon. Not. Roy. Astron. Soc.} {\bf 428} (2013) 3648,
  [\href{http://arxiv.org/abs/1210.7787}{{\tt arXiv:1210.7787}}].

\bibitem{Naoz:2014bqa}
S.~Naoz and R.~Narayan, {\it {Globular Clusters and Dark Satellite Galaxies
  through the Stream Velocity}},  {\em Astrophys. J.} {\bf 791} (2014) L8,
  [\href{http://arxiv.org/abs/1407.3795}{{\tt arXiv:1407.3795}}].

\bibitem{Popa:2015lkr}
C.~Popa, S.~Naoz, F.~Marinacci, and M.~Vogelsberger, {\it {Gas rich and gas
  poor structures through the stream velocity effect}},  {\em Mon. Not. Roy.
  Astron. Soc.} {\bf 460} (2016), no.~2 1625--1639,
  [\href{http://arxiv.org/abs/1512.06862}{{\tt arXiv:1512.06862}}].

\bibitem{Allsman:2000kg}
{\bf Macho} Collaboration, R.~A. Allsman et~al., {\it {MACHO project limits on
  black hole dark matter in the 1-30 solar mass range}},  {\em Astrophys. J.}
  {\bf 550} (2001) L169, [\href{http://arxiv.org/abs/astro-ph/0011506}{{\tt
  astro-ph/0011506}}].

\bibitem{Tisserand:2006zx}
{\bf EROS-2} Collaboration, P.~Tisserand et~al., {\it {Limits on the Macho
  Content of the Galactic Halo from the EROS-2 Survey of the Magellanic
  Clouds}},  {\em Astron. Astrophys.} {\bf 469} (2007) 387--404,
  [\href{http://arxiv.org/abs/astro-ph/0607207}{{\tt astro-ph/0607207}}].

\bibitem{Wyrzykowski:2011tr}
L.~Wyrzykowski et~al., {\it {The OGLE View of Microlensing towards the
  Magellanic Clouds. IV. OGLE-III SMC Data and Final Conclusions on MACHOs}},
  {\em Mon. Not. Roy. Astron. Soc.} {\bf 416} (2011) 2949,
  [\href{http://arxiv.org/abs/1106.2925}{{\tt arXiv:1106.2925}}].

\bibitem{Griest:2013aaa}
K.~Griest, A.~M. Cieplak, and M.~J. Lehner, {\it {Experimental Limits on
  Primordial Black Hole Dark Matter from the First 2 yr of Kepler Data}},  {\em
  Astrophys. J.} {\bf 786} (2014), no.~2 158,
  [\href{http://arxiv.org/abs/1307.5798}{{\tt arXiv:1307.5798}}].

\bibitem{Oguri:2017ock}
M.~Oguri, J.~M. Diego, N.~Kaiser, P.~L. Kelly, and T.~Broadhurst, {\it
  {Understanding caustic crossings in giant arcs: characteristic scales, event
  rates, and constraints on compact dark matter}},  {\em Phys. Rev.} {\bf D97}
  (2018), no.~2 023518, [\href{http://arxiv.org/abs/1710.00148}{{\tt
  arXiv:1710.00148}}].

\bibitem{Mediavilla:2017bok}
E.~Mediavilla, J.~Jiménez-Vicente, J.~A. Muñoz, H.~Vives-Arias, and
  J.~Calderón-Infante, {\it {Limits on the Mass and Abundance of Primordial
  Black Holes from Quasar Gravitational Microlensing}},  {\em Astrophys. J.}
  {\bf 836} (2017), no.~2 L18, [\href{http://arxiv.org/abs/1702.00947}{{\tt
  arXiv:1702.00947}}].

\bibitem{Zhang:2010qr}
C.~M. Zhang, J.~Wang, Y.~H. Zhao, H.~X. Yin, L.~M. Song, D.~P. Menezes, D.~T.
  Wickramasinghe, L.~Ferrario, and P.~Chardonnet, {\it {Study of measured
  pulsar masses and their possible conclusions}},  {\em Astron. Astrophys.}
  {\bf 527} (2011) A83, [\href{http://arxiv.org/abs/1010.5429}{{\tt
  arXiv:1010.5429}}].

\bibitem{Casares:2017jah}
J.~Casares, P.~G. Jonker, and G.~Israelian, {\it {X-ray Binaries}},
  \href{http://arxiv.org/abs/1701.07450}{{\tt arXiv:1701.07450}}.

\bibitem{Shakura:1972te}
N.~I. Shakura and R.~A. Sunyaev, {\it {Black holes in binary systems.
  Observational appearance}},  {\em Astron. Astrophys.} {\bf 24} (1973)
  337--355.

\bibitem{Rappaport:2003un}
S.~A. Rappaport, J.~M. Fregeau, and H.~Spruit, {\it {Accretion onto fast x-ray
  pulsars}},  {\em Astrophys. J.} {\bf 606} (2004) 436--443,
  [\href{http://arxiv.org/abs/astro-ph/0310224}{{\tt astro-ph/0310224}}].

\bibitem{Hickox:2004fy}
R.~C. Hickox, R.~Narayan, and T.~R. Kallman, {\it {Origin of the soft excess in
  x-ray pulsars}},  {\em Astrophys. J.} {\bf 614} (2004) 881--896,
  [\href{http://arxiv.org/abs/astro-ph/0407115}{{\tt astro-ph/0407115}}].

\bibitem{Hickox:2005nd}
R.~C. Hickox and S.~D. Vrtilek, {\it {Pulse-phase spectroscopy of SMC X-1 with
  Chandra and XMM-Newton: Reprocessing by a precessing disk?}},  {\em
  Astrophys. J.} {\bf 633} (2005) 1064--1075,
  [\href{http://arxiv.org/abs/astro-ph/0506438}{{\tt astro-ph/0506438}}].

\bibitem{Hung:2010cf}
L.-W. Hung, R.~C. Hickox, B.~Boroson, and S.~D. Vritlek, {\it {Suzaku X-ray
  Spectra and Pulse Profile Variations during the Superorbital Cycle of LMC
  X-4}},  {\em Astrophys. J.} {\bf 720} (2010) 1202--1214,
  [\href{http://arxiv.org/abs/1007.3280}{{\tt arXiv:1007.3280}}].

\bibitem{Graham:2006ae}
A.~W. Graham, D.~Merritt, B.~Moore, J.~Diemand, and B.~Terzic, {\it {Empirical
  Models for Dark Matter Halos. II. Inner profile slopes, dynamical profiles,
  and $\rho/\sigma^3$}},  {\em Astron. J.} {\bf 132} (2006) 2701--2710,
  [\href{http://arxiv.org/abs/astro-ph/0608613}{{\tt astro-ph/0608613}}].

\bibitem{Hilditch:2004pz}
R.~W. Hilditch, I.~D. Howarth, and T.~J. Harries, {\it {Forty eclipsing
  binaries in the Small Magellanic Cloud: Fundamental parameters and cloud
  distance}},  {\em Mon. Not. Roy. Astron. Soc.} {\bf 357} (2005) 304--324,
  [\href{http://arxiv.org/abs/astro-ph/0411672}{{\tt astro-ph/0411672}}].

\bibitem{Keller:2006ek}
S.~C. Keller and P.~R. Wood, {\it {Bump cepheids in the magellanic clouds:
  metallicities, the distances to the lmc and smc, and the pulsation-evolution
  mass discrepancy}},  {\em Astrophys. J.} {\bf 642} (2006) 834--841,
  [\href{http://arxiv.org/abs/astro-ph/0601225}{{\tt astro-ph/0601225}}].

\bibitem{LMC-X-4}
\url{http://simbad.u-strasbg.fr/simbad/sim-id?Ident=LMC%20X-4}.

\bibitem{x-ray-pulsars}
\url{https://www.sternwarte.uni-erlangen.de/wiki/index.php/List_of_accreting_X-ray_pulsars}.

\bibitem{Nesti:2013uwa}
F.~Nesti and P.~Salucci, {\it {The Dark Matter halo of the Milky Way, AD
  2013}},  {\em JCAP} {\bf 1307} (2013) 016,
  [\href{http://arxiv.org/abs/1304.5127}{{\tt arXiv:1304.5127}}].

\bibitem{Matsunaga:2006uc}
N.~Matsunaga and K.~Yamamoto, {\it {The finite source size effect and the wave
  optics in gravitational lensing}},  {\em JCAP} {\bf 0601} (2006) 023,
  [\href{http://arxiv.org/abs/astro-ph/0601701}{{\tt astro-ph/0601701}}].

\bibitem{Neilsen:2004eb}
J.~Neilsen, R.~C. Hickox, and S.~D. Vrtilek, {\it {Phase variation in the pulse
  profile of SMC X-1}},  {\em Astrophys. J.} {\bf 616} (2004) L135--L138,
  [\href{http://arxiv.org/abs/astro-ph/0410495}{{\tt astro-ph/0410495}}].

\bibitem{Goodman}
K.~Z. {Stanek}, B.~{Paczynski}, and J.~{Goodman}, {\it {Features in the spectra
  of gamma-ray bursts}},  {\em Astrophysical Journal Letters} {\bf 413} (Aug.,
  1993) L7--L10.

\bibitem{1991ApJ...372L..79G}
K.~Griest, C.~Alcock, T.~S. Axelrod, D.~P. Bennett, K.~H. Cook, K.~C. Freeman,
  H.-S. Park, S.~Perlmutter, B.~A. Peterson, P.~J. Quinn, A.~W. Rodgers, C.~W.
  Stubbs, and M.~Collaboration, {\it {Gravitational microlensing as a method of
  detecting disk dark matter and faint disk stars}},  {\em Astrophysical
  Journal Letters} {\bf 372} (May, 1991) L79--L82.

\bibitem{doi:10.1093/mnras/stw2775}
D.~R. Cole and J.~Binney, {\it A centrally heated dark halo for our galaxy},
  {\em Monthly Notices of the Royal Astronomical Society} {\bf 465} (2017),
  no.~1 798--810.

\bibitem{doi:10.1093/mnras/197.2.247}
M.~J. Coe, S.~J.~B. Burnell, A.~R. Engel, A.~J. Evans, and J.~J. Quenby, {\it
  The x-ray spectrum of smc x-1 observed from the ariel v satellite},  {\em
  Monthly Notices of the Royal Astronomical Society} {\bf 197} (1981), no.~2
  247--251.

\bibitem{1991ApJ...381..101L}
A.~{Levine}, S.~{Rappaport}, A.~{Putney}, R.~{Corbet}, and F.~{Nagase}, {\it
  {LMC X-4 - GINGA observations and search for orbital period changes}},  {\em
  Astrophys. J.} {\bf 381} (Nov., 1991) 101--109.

\bibitem{doi:10.1117/12.2062667}
K.~P. Singh, S.~N. Tandon, P.~C. Agrawal, and et. al., {\it Astrosat mission},
  {\em Proc.SPIE} {\bf 9144} (2014) 9144--9144--15.

\bibitem{RXTE-effective-area}
\url{https://heasarc.gsfc.nasa.gov/docs/xte/RXTE_tech_append.pdf}.

\bibitem{RXTE_Heasoft}
\url{https://heasarc.gsfc.nasa.gov/docs/xte/xhp_proc_analysis.html}.

\bibitem{Heasoft}
\url{https://heasarc.gsfc.nasa.gov/docs/software/lheasoft/}.

\bibitem{Griest:2011av}
K.~Griest, M.~J. Lehner, A.~M. Cieplak, and B.~Jain, {\it {Microlensing of
  Kepler Stars as a Method of Detecting Primordial Black Hole Dark Matter}},
  {\em Phys. Rev. Lett.} {\bf 107} (2011) 231101,
  [\href{http://arxiv.org/abs/1109.4975}{{\tt arXiv:1109.4975}}].

\bibitem{Raichur:2009ej}
H.~Raichur and B.~Paul, {\it {Effect of pulse profile variations on measurement
  of eccentricity in orbits of Cen X-3 and SMC X-1}},  {\em Mon. Not. Roy.
  Astron. Soc.} {\bf 401} (2010) 1532,
  [\href{http://arxiv.org/abs/0909.4271}{{\tt arXiv:0909.4271}}].

\bibitem{Rai:2018vkw}
B.~Rai, P.~Pradhan, and B.~C. Paul, {\it {A report on the type II X-ray burst
  from SMC X-1}},  \href{http://arxiv.org/abs/1806.03244}{{\tt
  arXiv:1806.03244}}.

\bibitem{Heasarc_browse}
\url{https://heasarc.gsfc.nasa.gov/cgi-bin/W3Browse/w3browse.pl}.

\bibitem{Athena}
X.~Barcons, K.~Nandra, D.~Barret, J.-W. den Herder, A.~C. Fabian, L.~Piro,
  M.~G. Watson, and the Athena~team, {\it Athena: the x-ray observatory to
  study the hot and energetic universe},  {\em Journal of Physics: Conference
  Series} {\bf 610} (2015), no.~1 012008.

\bibitem{LynxTeam:2018usc}
{\bf Lynx Team} Collaboration, {\it {The Lynx Mission Concept Study Interim
  Report}},  \href{http://arxiv.org/abs/1809.09642}{{\tt arXiv:1809.09642}}.

\bibitem{Zhang:2016ach}
{\bf eXTP} Collaboration, S.~N. Zhang et~al., {\it {eXTP -- enhanced X-ray
  Timing and Polarimetry Mission}},  {\em Proc. SPIE Int. Soc. Opt. Eng.} {\bf
  9905} (2016) 99051Q, [\href{http://arxiv.org/abs/1607.08823}{{\tt
  arXiv:1607.08823}}].

\bibitem{Feroci:2011jc}
{\bf LOFT} Collaboration, M.~Feroci et~al., {\it {The Large Observatory for
  X-ray Timing (LOFT)}},  {\em Exper. Astron.} {\bf 34} (2012) 415,
  [\href{http://arxiv.org/abs/1107.0436}{{\tt arXiv:1107.0436}}].

\bibitem{Vacchi:2018mnt}
{\bf LOFT} Collaboration, A.~Vacchi, {\it {The LOFT mission concept}},  {\em
  Nucl. Part. Phys. Proc.} {\bf 297-299} (2018) 194--206.

\bibitem{Hu:2013wza}
C.-P. Hu, Y.~Chou, T.-C. Yang, and Y.-H. Su, {\it {Superorbital Phase-Resolved
  Analysis of SMC X-1}},  {\em Astrophys. J.} {\bf 773} (2013) 58,
  [\href{http://arxiv.org/abs/1306.5819}{{\tt arXiv:1306.5819}}].

\bibitem{Trowbridge:2007kj}
S.~Trowbridge, M.~A. Nowak, and J.~Wilms, {\it {Tracking the Orbital and
  Super-orbital Periods of SMC X-1}},  {\em Astrophys. J.} {\bf 670} (2007)
  624, [\href{http://arxiv.org/abs/0708.0038}{{\tt arXiv:0708.0038}}].

\bibitem{Nowak:2000kf}
M.~A. Nowak, J.~Wilms, W.~A. Heindl, K.~Pottschmidt, J.~B. Dove, and M.~C.
  Begelman, {\it {A good long look at the black hole candidates lmc x-1 and lmc
  x-3}},  {\em Mon. Not. Roy. Astron. Soc.} {\bf 320} (2001) 316,
  [\href{http://arxiv.org/abs/astro-ph/0005487}{{\tt astro-ph/0005487}}].

\bibitem{Refsdal:1993kf}
S.~Refsdal and J.~Surdej, {\it {Gravitational lenses}},  {\em Rept. Prog.
  Phys.} {\bf 57} (1994) 117--186.

\bibitem{Gould:1992yv}
A.~Gould, {\it {Extending the MACHO search to about 10 exp 6 solar masses}},
  {\em Astrophys. J.} {\bf 392} (1992) 442--451.

\bibitem{Gould:1993yv}
A.~Gould, {\it {Proper motions of MACHOs}},  {\em Astrophys. J.} {\bf 421}
  (1994) L71--L74.

\bibitem{Foley:2016obj}
R.~J. Foley et~al., {\it {Ultraviolet Diversity of Type Ia Supernovae}},  {\em
  Mon. Not. Roy. Astron. Soc.} {\bf 461} (2016), no.~2 1308--1316,
  [\href{http://arxiv.org/abs/1604.01021}{{\tt arXiv:1604.01021}}].

\bibitem{white}
{Werner, K.} and {Rauch, T.}, {\it Analysis of hst/cos spectra of the bare c-o
  stellar core h1504+65 and a high-velocity twin in the galactic halo},  {\em
  Astronomy and astrophysics} {\bf 584} (2015) A19.

\bibitem{Bai:2018dxf}
Y.~Bai, A.~J. Long, and S.~Lu, {\it {Dark Quark Nuggets}},
  \href{http://arxiv.org/abs/1810.04360}{{\tt arXiv:1810.04360}}.

\bibitem{Kolb:1993zz}
E.~W. Kolb and I.~I. Tkachev, {\it {Axion miniclusters and Bose stars}},  {\em
  Phys. Rev. Lett.} {\bf 71} (1993) 3051--3054,
  [\href{http://arxiv.org/abs/hep-ph/9303313}{{\tt hep-ph/9303313}}].

\end{thebibliography}\endgroup

\end{document}